\documentclass[]{spie}  %>>> use for US letter paper
\usepackage{amsmath,amsfonts,amssymb}
\usepackage{graphicx}
\usepackage[colorlinks=true, allcolors=blue]{hyperref}
\usepackage{subfig}
\usepackage{float}

\title{The Santa Cruz Extreme AO Lab (SEAL) 2.0: A reflective, multi-wavelength rebuild}

\author[a]{Rebecca Jensen-Clem}
\author[b]{Vincent Chambouleyron}
\author[a]{Prince Javier}
\author[c]{Daren Dillon}
\author[a]{Emiel H. Por}
\author[d]{Benjamin Calvin}
\author[e]{Sylvain Cetre}
\author[f]{Rodrigo Amezcua Correa}
\author[f]{Tara Crowe}
\author[a]{Jordan Diaz}
\author[f]{Caleb Dobias}
\author[g,h]{David Doelman}
\author[f]{Stephen Eikenberry}
\author[a]{J. Fowler}
\author[i]{Benjamin L. Gerard}
\author[a,c]{Phil Hinz}
\author[c]{Renate Kupke}
\author[a]{Ashai Moreno}
\author[a]{Tiffany Nguyen}
\author[a]{Maissa Salama}
\author[a]{Aditya R. Sengupta}
\author[j]{Nour Skaf}
\author[g]{Frans Snik}
\affil[a]{Univ. of California, Santa Cruz (United States)}
\affil[b]{Laboratoire d'Astrophysique de Marseille (France)}
\affil[c]{Univ. of California Observatories (United States)}
\affil[d]{Univ. of California, Los Angeles (United States)}
\affil[e]{Wakea Consulting (France)}
\affil[f]{University of Central Florida (United States)}
\affil[g]{Leiden Observatory (Netherlands)}
\affil[h]{SRON Netherlands Institute for Space Research (Netherlands)}
\affil[i]{Lawrence Livermore National Laboratory (United States)}
\affil[j]{Institute for Astronomy, University of Hawai`i, 640 N. Aohoku Place, Hilo, HI, 96720, United States}

\authorinfo{Further author information: (Send correspondence to R.J.-C.)\\R.J.-C.: E-mail: rjensenc@ucsc.edu.edu}

\begin{document} 
\maketitle

\begin{abstract}
The Santa cruz Extreme Adaptive optics Lab (SEAL) is a visible/near-infrared wavelength testbed designed to support technology development for high contrast imaging on large, segmented, ground-based telescopes. SEAL saw first light in 2021 as a transmissive, visible-wavelength AO testbed. In this paper, we present four major upgrades to SEAL: (1) the testbed has been rebuilt with custom off-axis parabolic mirrors, enabling operation in both near-infrared and visible wavelengths; (2) the suite of wavefront sensors now includes a Shack-Hartmann, transmissive four-sided pyramid, vector-Zernike, and, in the muirSEAL testbed, a photonic lantern; (3) the testbed includes a vector-vortex coronagraph and will soon include a hybrid astrophotonic coronagraph; (4) in addition to its original Keck-heritage RTC, SEAL now includes two additional control software packages: Catkit, originally developed for the HiCAT testbed at the Space Telescope Science Institute, and the RTC Compute And Control for Adaptive Optics (CACAO), originally designed for Subaru/SCExAO. We discuss the performance of the testbed after the reflective rebuild and on-going technology development work at SEAL. 
\end{abstract}


\section{INTRODUCTION}
\label{sec:intro}

The Santa cruz Extreme Adaptive optics Lab (SEAL) was originally introduced in Jensen-Clem et al.~2021\cite{seal2021} as a transmissive, visible-wavelength testbed aimed at high contrast technology development for segmened, ground-based telescopes. In 2021, SEAL included a spatial light modulator (SLM) to simulate atmospheric turbulence, a segmented deformable mirror (DM) to simulate the W.~M.~Keck Observatory (WMKO) segmented primary mirror, a woofer/tweeter DM system for wavefront correction, four wavefront sensor arms (Shack-Hartmann, reflective three-sided pyramid, vector-Zernike, and Fast Atmospheric Self-Coherent Camera), and a real time control (RTC) system identical to the Keck II Telescope's pyramid wavefront sensor testbed\cite{Cetre2018}. While some aspects of the testbed are relatively unchanged (e.g. the SLM and DMs), others have been replaced, rebuilt, and extended (e.g. the wavefront sensing and software capabilities). Furthermore, SEAL was completely rebuilt in the summer and fall of 2024 with custom off-axis parabolic mirrors, enabling operation in both near-infrared and visible wavelengths. 

The remainder of this paper focuses on the research and upgrades to SEAL since 2021, with an emphasis on the recent reflective, multi-wavelength rebuild. Section \ref{sec:rebuild} describes the rebuild itself; Section \ref{sec:performance} describes the post-rebuild performance; Section \ref{sec:software} describes the next-generation of SEAL software; Section \ref{sec:current} describes on-going and future AO and coronagraphy work at SEAL.

\subsection{SUMMARY OF SEAL WORK TO DATE}

Since SEAL was introduced in Jensen-Clem et al.~2021\cite{seal2021}, it has been used to develop a variety of wavefront sensing and control technologies, including: 
\begin{itemize}
    \item \textbf{Zernike Wavefront Sensing:} Salama et al.~2022\cite{salama2022} presented the design and performance of a vector Zernike wavefront sensor (vZWFS; also discussed in Sections \ref{sec:performance} and \ref{sec:current} of this paper), which has been used to reconstruct IrisAO piston errors down to $1\,$nm. Chambouleyron et al.~2024\cite{Chambouleyron2024} used SEAL's vZWFS and IrisAO DM to test non-linear methods of Zernike wavefront reconstruction including the arcsin, phase-shifted ZWFS (PSZWFS) arcsin, and PSZWFS Gerchberg-Saxton reconstructors, concluding that the combination of the PSZWFS and non-linear wavefront reconstruction improved the dynamic range compared with the classical ZWFS. The testbed was additionally employed to evaluate a reflective ZWFS in preparation for the GPI2.0 instrument \cite{Chambouleyron2023_SPIE}, which will incorporate this sensor for non-common path aberration (NCPA) measurements.

    \item \textbf{Pyramid Wavefront Sensing:} Gerard et al.~2021\cite{bright2021} introduced the concept of the bright PWFS (bPWFS) in which a piston phase offset is added to a region of radius $0.5\, \lambda/$D around the tip of a standard PWFS. The goal of the bPWFS is to improve the measurement error and linearity compared with the standard PWFS. Gerard et al.~2022\cite{various2022} tested the non-modulated bPWFS on SEAL, showing improved linearity compared with the non-modulated standard PWFS. Chambouleyron et al. 2024\cite{2024GS_SEAL} demonstrated the application of the Gerchberg–Saxton algorithm for non-linear reconstruction of non-modulated pyramid wavefront sensor signals, resulting in a substantial enhancement of the sensor’s capture range.

    \item \textbf{Chopper-based Wavefront Sensing:} Gerard et al.~2022\cite{various2022} discussed two methods of chopper-based wavefront sensing: (1) using an optical chopper to partially block and unblock pupil frames while recording simultaneous focal plane images -- this technique was demonstrated on SEAL at high speed in Gerard et al.~2023\cite{Gerard2023chop}; (2) instead of using an external chopper, using the AO system's existing DM to apply a local tilt to a fraction of the pupil -- this technique was demonstrated on SEAL in Soto et al.~2023\cite{Soto2023}, showing that as little  as $250\,$nm of DM stroke is needed to create a linear signal for low-order Zernike modes. 

    \item \textbf{Fast Atmospheric Self coherent camera Technique (FAST):} FAST\cite{2018AJ....156..106G,2020PhDT........17G,Gerard2021} is a focal plane wavefront sensing technique that uses the interference of coherent starlight to sense the speckle electric field in the coronagraphic image. Gerard et al.~2022\cite{Gerard2022} used SEAL's FAST setup\cite{10.1117/12.2599556} to demonstrate an up to $5\times$ contrast improvement as FAST compensated for evolving residual atmospheric turbulence in closed loop. Gerard et al.~2023\cite{Gerard2023} went on to demonstrate simultaneous first and second stage AO control using SEAL's SHWFS and FAST sensors at up to $200\,$Hz loop speeds. This work has led to a current project to deploy FAST on-sky at the Shane Telescope at Lick Observatory\cite{Gerard2025}. Additionally, the FAST setup was used to demonstrate linear-quadratic-gaussian (LQG) Control for tip/tilt: Sengupta et al.~2022\cite{Sengupta2022} showed that LQG control with FAST improved the RMS wavefront error over all temporal frequencies (compared with an integrator) and demonstrated frequency-domain notching of an injected $9\,$Hz vibration mode.

    \item \textbf{Simultaneous Segmented Primary and AO Control:} Calvin et al.~2024\cite{Calvin2024} and Calvin et al~2025 in these proceedings used SEAL to experimentally validate a technique\cite{Calvin2025a,Calvin2025b} for projecting the wavefront reconstructed by a SHWFS onto the space controllable by a telescope's segmented primary mirror and by an AO system's DM. Calvin et al~2025 in these proceedings simulated atmospheric turbulence using SEAL's BMC DM, a segmented primary using the IrisAO DM, and an AO system DM using the ALPAO DM. The SHWFS was then used to control both the IrisAO and ALPAO DMs in closed-loop, successfully improving the Strehl ratio by simultaneously controlling co-phasing errors and atmospheric turbulence. 

\end{itemize}

In addition to these major wavefront sensing projects, the SEAL testbed continued to be built out and characterized as planned in Jensen-Clem et al.~2021\cite{seal2021}. First, van Kooten et al.~2022\cite{Kooten2022} characterized SEAL's spatial light modulator (SLM), showing that single layer turbulence with $r_0 = 15\,$cm applied to the SLM produced a long-exposure PSF with the corresponding FWHM$=0.98 \lambda/r_0$.  van Kooten et al.~2022 additionally showed that the SLM can run as fast as $250\,$Hz. Second, Moreno et al.~2024\cite{Moreno2024} installed and characterized SEAL's vector vortex coronagraph arm, finding an average raw contrast of $2\times 10^{-4}$ between 2 and 4 $\lambda /$D. This raw contrast is about $100\times$ higher than expected based on simulations, suggesting that the performance is currently limited by the vortex mask leakage term and polarizer efficiency. The next generation of coronagraphy with SEAL will be astrophotonics-based, as described in Section \ref{sec:current}.

\section{REFLECTIVE REBUILD}
\label{sec:rebuild}
\begin{figure} [H]
    \centering
    \subfloat[]{\includegraphics[width=0.5\linewidth]{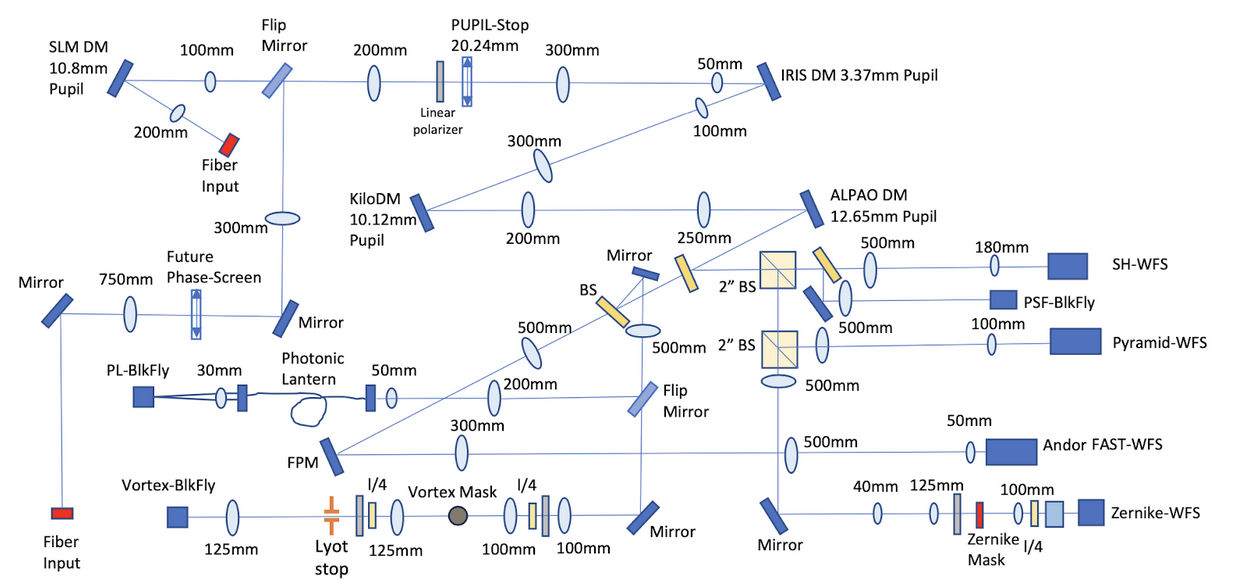
    }\label{fig:SEALOG}}
    \hfill
    \subfloat[]{\includegraphics[width=0.5\linewidth]{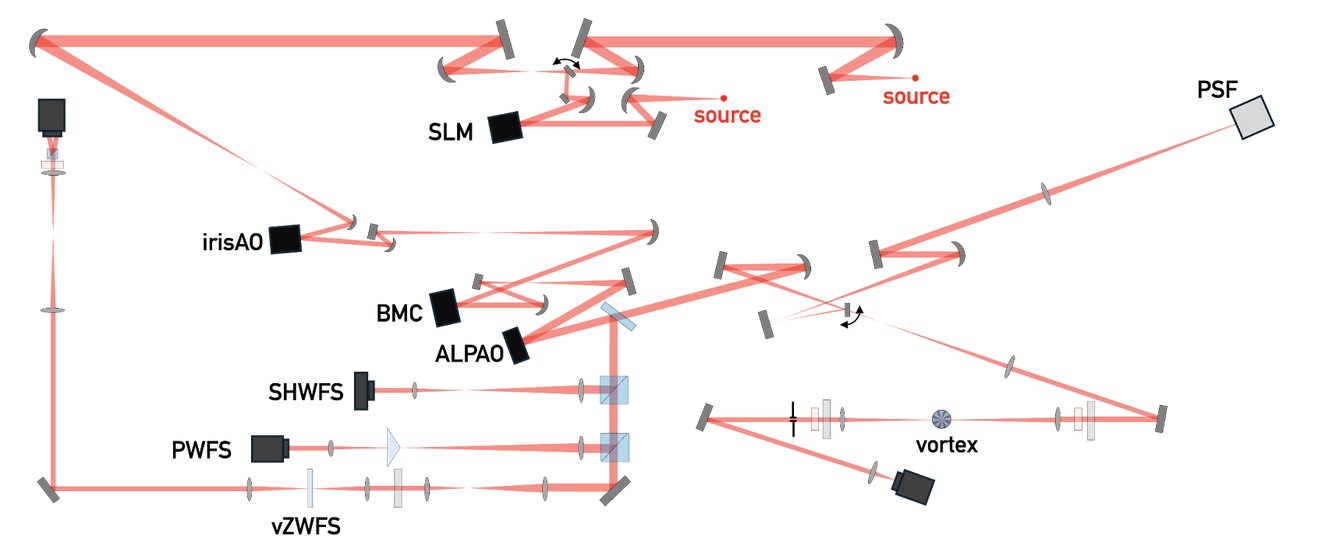}\label{fig:SEALNEW}}
    \caption{(a) Schematic of the refractive SEAL testbed (b) Schematic of the updated reflective SEAL testbed with lenses replaced by OAPs. Light from the fiber input sources are relayed through the DMs (IrisAO, BMC, and ALPAO), into the wavefront sensors and coronagraphs.}
\end{figure}

In order to feed the testbed's array of wavefront sensors and cornographs with well-corrected light at both near-infrared and visible wavelengths, SEAL's refractive lenses (see Figure~\ref{fig:SEALOG}) were replaced with 13 reflective custom off-axis parabolic mirrors (OAPs), while maintaining the architecture of the original system. The set of 13 OAPs relay pupils between the simulated Keck telescope aperture from the 37-segment IrisAO DM and the subsequent DMs, into the series of wavefront sensors and coronagraphs placed after each of the DMs to be used in a closed loop (see Figure~\ref{fig:SEALNEW}).

The upgrade from lenses to OAPs provides near on-axis performance, an increase in optical throughput, elimination of chromatic aberrations, and decreased overall mechanical size, allowing more space for the testbed's wavefront sensors and coronagraphs. Each of the 13 OAPs manufactured for the SEAL testbed have a specified off-axis angle (OAA), off-axis distance (OAD), focal length, and diameter, which differ to varying degrees from their as-built parameters due to manufacturing errors. Up to 100 $\mu m$ errors were accepted for the focal length and OAD as per the tolerance analysis done for each of the OAPs, but errors that surpassed this set tolerance were implemented into the as-built design. To mitigate discrepancies between the Zemax optical design and the as-built testbed, the OAA, OAD, focal length, and diameter of each OAPs were characterized.

\begin{figure} [h!]
    \centering
    \includegraphics[width=0.7\linewidth]{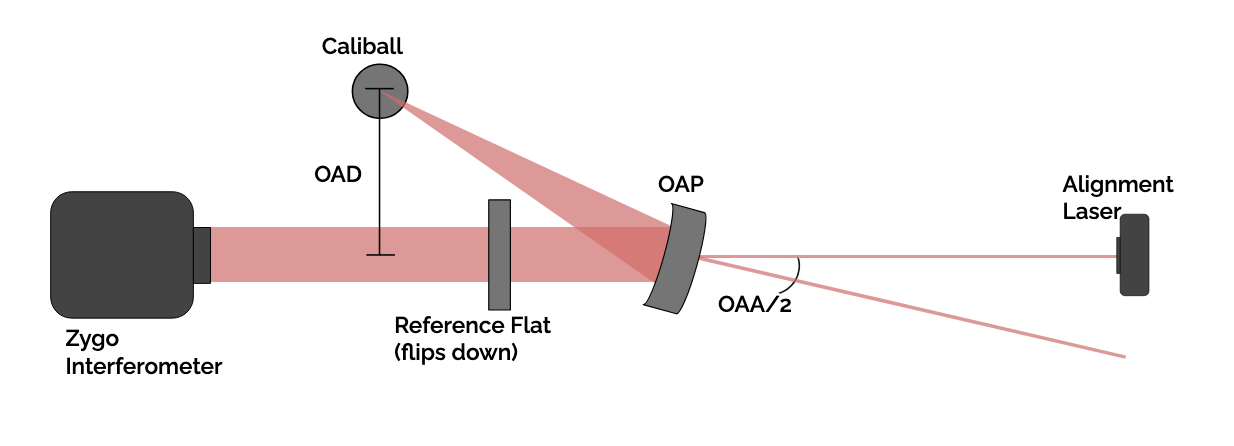}
    \caption{Schematic diagram of the OAP testing setup. The collimated beam from the Zygo Interferometer, which is aligned using the reference flat, is incident upon the OAP to be tested, such that the beam is focused at a specified distance and angle away from the optical axis. The caliball, placed at the focal point, acts as a point source that reflects back off the OAP and into the interferometer. The resulting interference fringes are read out along with the rms wavefront error on the interferometer.}
    \label{fig:TestingSchematic}
\end{figure}

The focal length, OAD, and OAA were verified using a Zygo interferometer. The testing shematic is shown in Figure~\ref{fig:TestingSchematic}. A majority of the OAPs surpassed the accepted 100 $\mu m$ tolerance in both focal length and OAD, deviating from their specified parameters on the order of millimeters.

After the OAPs were tested, each of the mirrors were mounted and adjusted for height - such that the incoming beam was concentric with the face of the OAP - and clocking - such that the Zernike term for the 45$^{\circ}$ astigmatism read out on a Shack-Hartmann wavefront sensor was at or close to zero. With the height and clocking set, the OAPs were placed sequentially in their approximate positions as dictated by the original Zemax design, and then shifted and rotated to accommodate the measured manufacturing errors. Although these shifts were large, the tolerances in the original design were such that the pupil sizes at the DM surfaces were still acceptable. After each subsequent OAP was placed in the system, we observed the wavefront - typically in collimated space - with a Shack-Hartmann wavefront sensor and made further adjustments to minimize defocus, 0$^{\circ}$/90$^{\circ}$ astigmatism, horizontal and vertical coma, and horizontal and vertical trefoil, as these are the main contributors to the rms wavefront error.

\section{ARCHITECTURE AND POST-REBUILD PERFORMANCE}
\label{sec:performance}

\subsection{Visible Wavefront Sensing Paths}

\begin{figure} [h!]
    \centering
\includegraphics[width=0.5\linewidth]{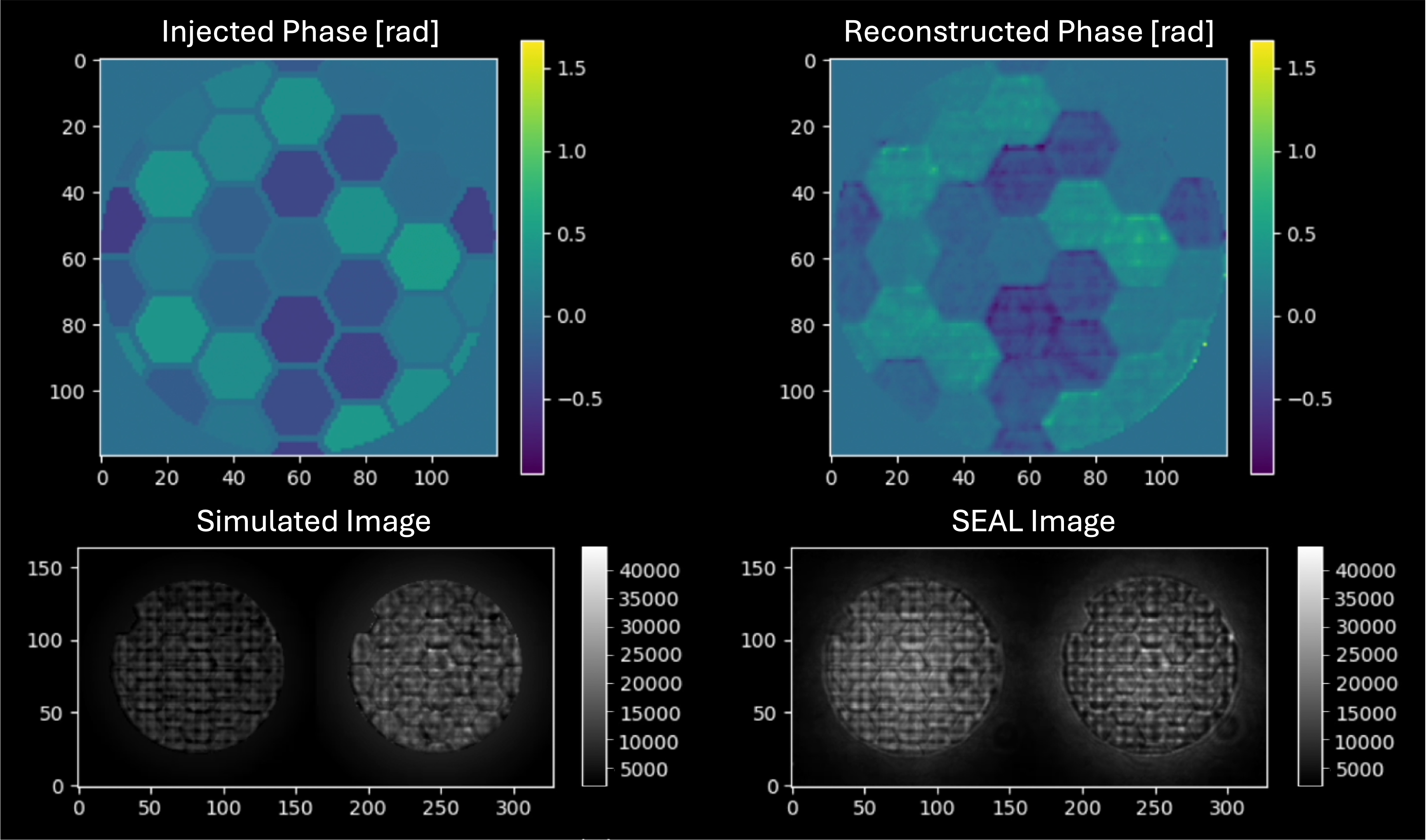}
    \caption{An example of the Zernike WFS closed-loop formance. Upper panels: the injected (piston-only) phase on the IrisAO segmented DM and the phase reconstructed by the Zernike WFS. Lower panels: the simulated and experimentally measured SEAL Zernike WFS images.}
    \label{fig:zernike_example}
\end{figure}

\begin{figure} [h!]
    \centering
    \includegraphics[width=0.6\linewidth]{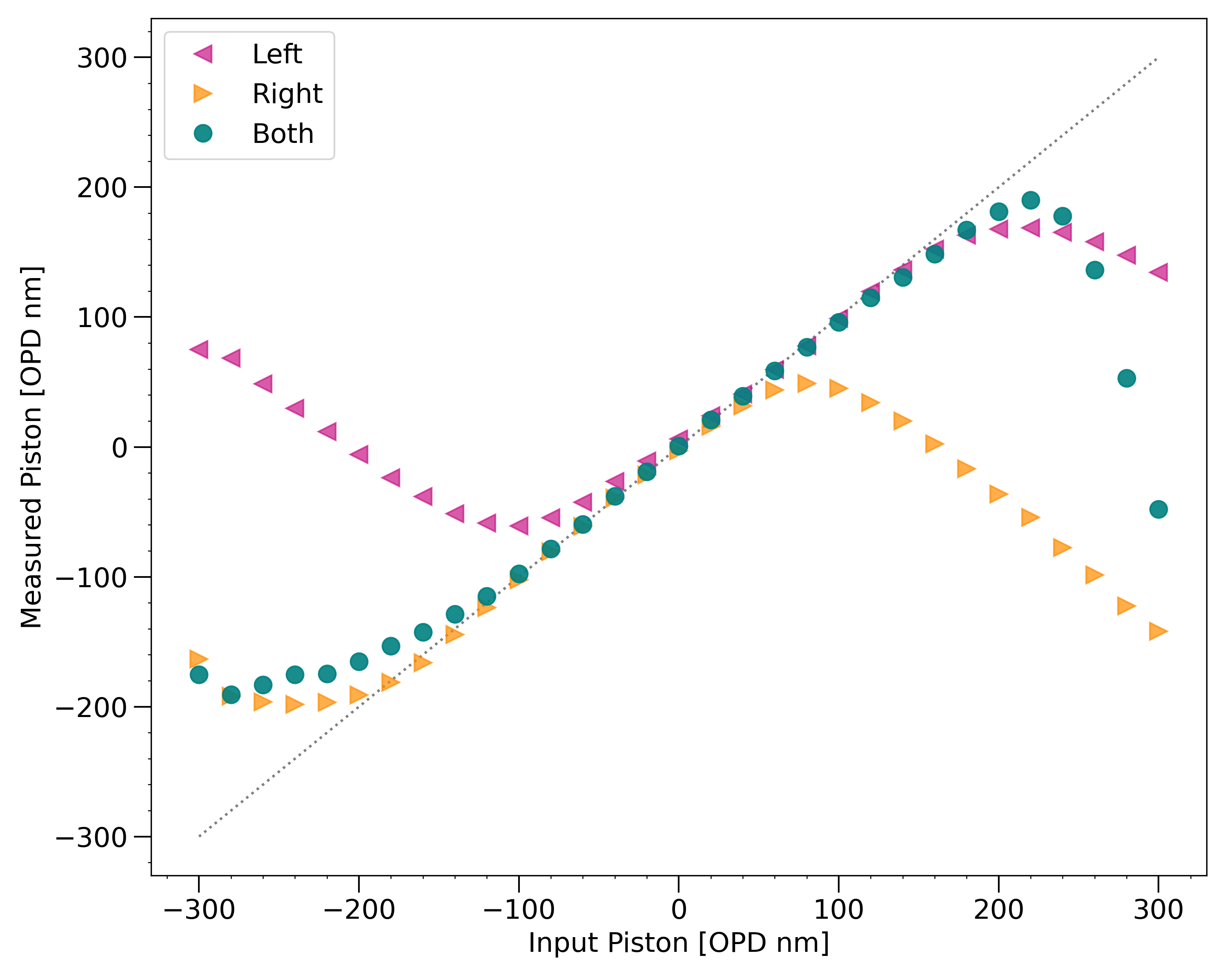}
    \caption{Zernike wavefront sensor dynamic range measured by poking in piston the central segment on the IRIS AO segmented DM. The pink left triangles show the reconstructed poke from using only the signal in the left pupil, the orange right triangles correspond to only using the right pupil, and the teal circles correspond to using both pupils for the phase reconstruction. The region where the response curve monotonically increases, nearly doubles when using both pupils. This demonstrates that using both pupils extends the dynamic range to $\lambda$ (or $2\pi$).}
    \label{fig:ZWFS_DynamicRange}
\end{figure}

As in the refractive SEAL testbed, all wavefront sensors currently operate in the visible (see Figure \ref{fig:SEALNEW}). These sensors are located after a beamsplitter placed immediately downstream of the ALPAO DM, and include a Shack-Hartmann WFS (SHWFS), non-modulated Pyramid WFS (PWFS), and vector Zernike WFS. Several modifications were introduced relative to the previous Figure \ref{fig:SEALOG} configuration:  

\begin{itemize}
    \item \textbf{Non-modulated PWFS branch:} The 3-sided reflective PWFS described in Jensen-Clem et al.~2021\cite{seal2021} was replaced with a double rooftop, 4-sided PWFS, following the design of the PWFS on SCExAO\cite{lozi2022}. As part of the reflective rebuild, the optical relay was modified to reduce the pupil image size on the detector, now spanning 106 pixels across the pupil diameter (compared to $\sim$400 pixels in the refractive SEAL layout). This reduction enables higher frame rates by exploiting the region-of-interest mode of the BlackFly camera, reaching up to 200~Hz. Despite the smaller pupil sampling, the system remains significantly oversampled relative to the highest controllable spatial frequency (BMC DM with 20 actuators across the pupil), thus allowing accurate measurement of fine phase aberrations on the bench.  

    \item \textbf{Zernike branch:} The Zernike branch was originally described in Salama et al.~2022\cite{salama2022}. The relay was re-designed to accommodate the liquid-crystal focal-plane mask that can be configured with diameters of 0.5~$\lambda/D$, 1~$\lambda/D$, or 2~$\lambda/D$ (at 635~nm). The Zernike branch is primarily used to develop segment co-phasing techniques in conjunction with the segmented IrisAO DM (see Section \ref{sec:current}). The closed-loop performance is illustrated in Figure \ref{fig:zernike_example}. The two pupil images in each of the lower panels of Figure \ref{fig:zernike_example} are due to the vector Zernike design first proposed by Doelman et al.~2019 \cite{doelman2019}. The advantage of the vector Zernike WFS design is illustrated in Figure \ref{fig:ZWFS_DynamicRange}: the Zernike WFS dynamic range is larger when measured using both pupils compared with using the left or the right pupil alone.
\end{itemize}

\begin{figure} [h!]
    \centering
\includegraphics[width=0.45\linewidth]{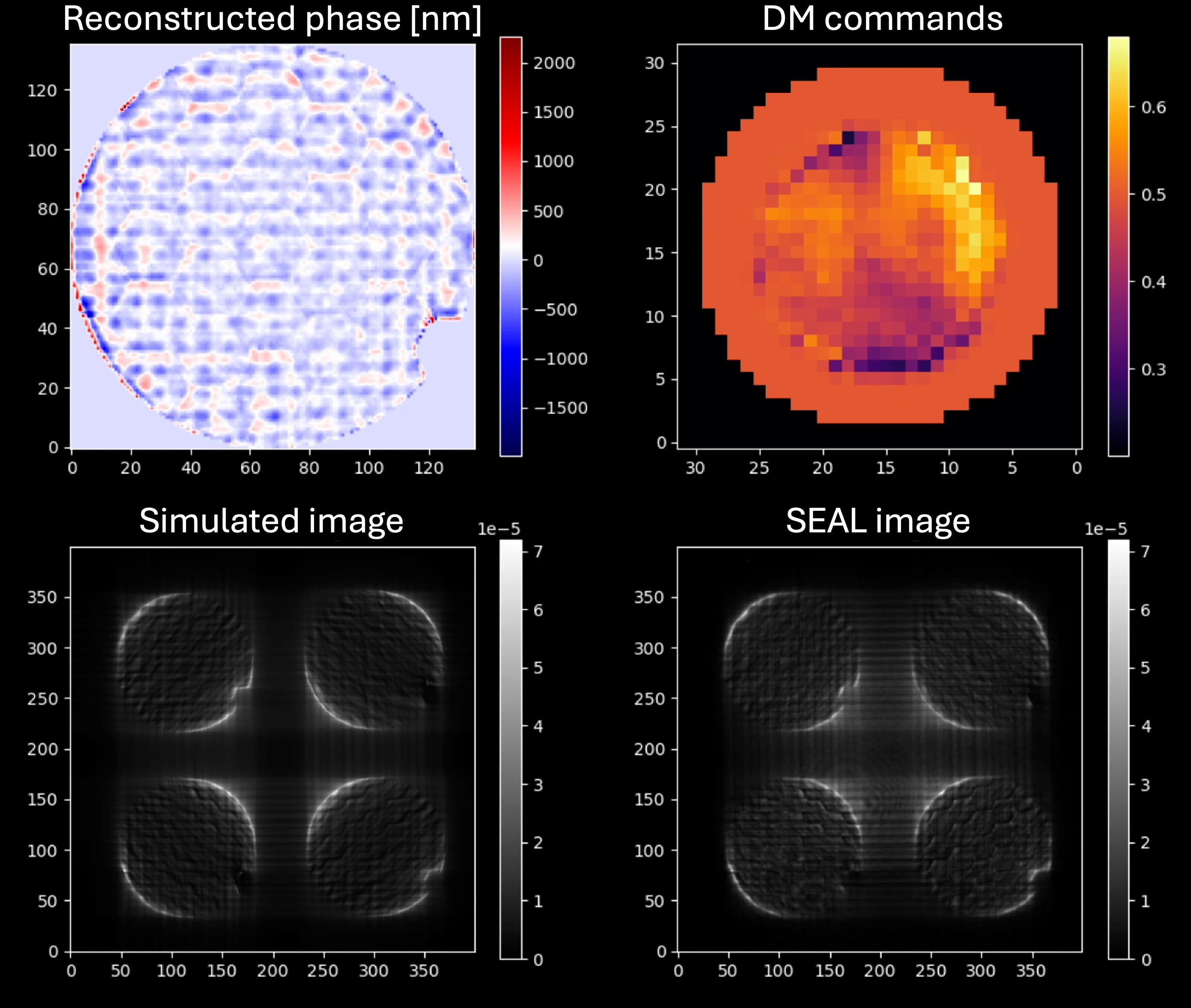}
    \caption{An example of the PWFS closed-loop performance. Top-left panel: the reconstructed phase from the PWFS, where the quilting pattern of the BMC DM and a dead segment from the IrisAO DM (on the lower right of the reconstructed phase panel) are visible. Top-right panel: Commands applied on BMC DM, where the SEAL aperture footprint on the DM is clearly visible. Bottom panels: the simulated PWFS signals (from the GS reconstruction) and the experimentally measured image.}
    \label{fig:pyarmid_example}
\end{figure}

\noindent \textbf{Startup procedure:} The testbed's static aberrations are mitigated using the non-modulated PWFS, simultaneously controlling the ALPAO DM and the BMC DM. The procedure is as follows: the ALPAO DM is initialized with its previously determined flat command, while the BMC DM is set to its mid-range position. The PWFS phase signal is reconstructed via a Gerchberg--Saxton algorithm (five iterations per reconstruction - see Chambouleyron et al. 2024\cite{2024GS_SEAL} for more details). Closed-loop correction is first performed with the ALPAO DM (larger stroke, coarser pitch) and subsequently refined with the BMC DM, yielding convergence towards a flat wavefront. The achieved residual error is $\sim$18~nm rms, comparable to the previous refractive SEAL results. The dominant contributors to the residual error after closed-loop correction are most likely the quilting effect of the BMC deformable mirror (clearly visible in the reconstructed phase shown in Figure \ref{fig:pyarmid_example}) and the small tilts present across the individual segments of the IrisAO DM. Figure~\ref{fig:pyarmid_example} illustrates the closed-loop performance.

\subsection{Visible Science Paths}

Two visible science channels are available, selectable via a pick-off mirror (mutually exclusive operation):  

\begin{itemize}
    \item \textbf{PSF imaging path:} A direct imaging channel equipped with an Andor Zyla 5.5 sCMOS camera.  
    \item \textbf{Coronagraphic path:} A visible vector vortex coronagraph, previously implemented as the visible coronagraph of the refractive SEAL testbed~\cite{Moreno2024}.  
\end{itemize}

\subsection{Infrared Science Path}

An infrared PSF imaging channel has been implemented. At present, illumination is switched between a visible laser (635~nm) and an infrared laser (1550~nm). The optical design for this branch is as follows: the collimated beam in the science PSF path is picked off by a mirror; a second mirror directs the collimated beam to an infrared lens (f = 250 mm), which converges the light onto the Goldeye G-130 GigE TEC1 infrared detector at the end of the branch. The infrared branch is aligned using a visible source to follow and manipulate the light’s path. This is possible due to the wavelength range of the infrared path's detector (\(\approx\) 400 to 1700 nm). To correct for wavefront aberrations, we close the loop on the PWFS using a visible source and ensure that there are no instabilities before introducing the infrared source. 

With the current setup, the measured Strehl ratio (SR) in the IR branch reaches 98\% after closing the loop with the visible PWFS, corresponding to an estimated NCPA of $\sim$30~nm rms between the IR PSF and the visible PWFS~\cite{asahi_thesis}. To estimate the NCPA, the Marechal approximation and the measured SR allowed for an estimate of the optical path difference (OPD) in the IR branch. Another OPD was then estimated for the PWFS branch using phase and pupil measurements from the wavefront sensor after closing the loop. With the two estimated OPDs, error propagation and the principles of adding normally distributed random variables were applied for the final NCPA estimate between the IR PSF and visible PWFS. 

Future upgrades to the IR branch will include the integration of a broadband white-light source and a dichroic beamsplitter. 

\newpage
\section{SOFTWARE REDESIGN}
\label{sec:software}

\subsection{Catkit2 as a bench control software}
The Catkit2 software framework\footnote{\url{https://github.com/spacetelescope/catkit2}} was originally developed for the HiCAT testbed at the Space Telescope Science Institute and provides a service-oriented architecture for laboratory control\cite{por_2024_11395554}. In this framework, hardware components and higher-level functions (e.g., control loops or safety systems) are represented as \emph{Services}. Each service, written in either Python or C/C++, runs in a separate process, enabling concurrent operation and reducing the risk of bottlenecks. Services expose an application programming interface (API) for parameter management, method execution, and low-latency data channels, such as streaming images or sending deformable mirror commands. A central Testbed process manages service lifecycle, configuration, and discovery. Client applications written in Python or C/C++ interact with services through proxy objects that abstract away the underlying inter-process communication.

While Catkit2 was originally developed for use on testbeds that use a single computer for control, SEAL instead uses a distributed hardware architecture composed of three computers. A primary Linux machine performs real-time control of deformable mirrors and wavefront sensor cameras. An additional two Windows machines operate peripheral devices, including light sources, a Meadowlark SLM, and an IrisAO segmented DM. This distribution of hardware between different machines requires reliable and low-latency communication between these nodes.

\begin{figure}
    \centering
    \includegraphics[width=0.95\columnwidth]{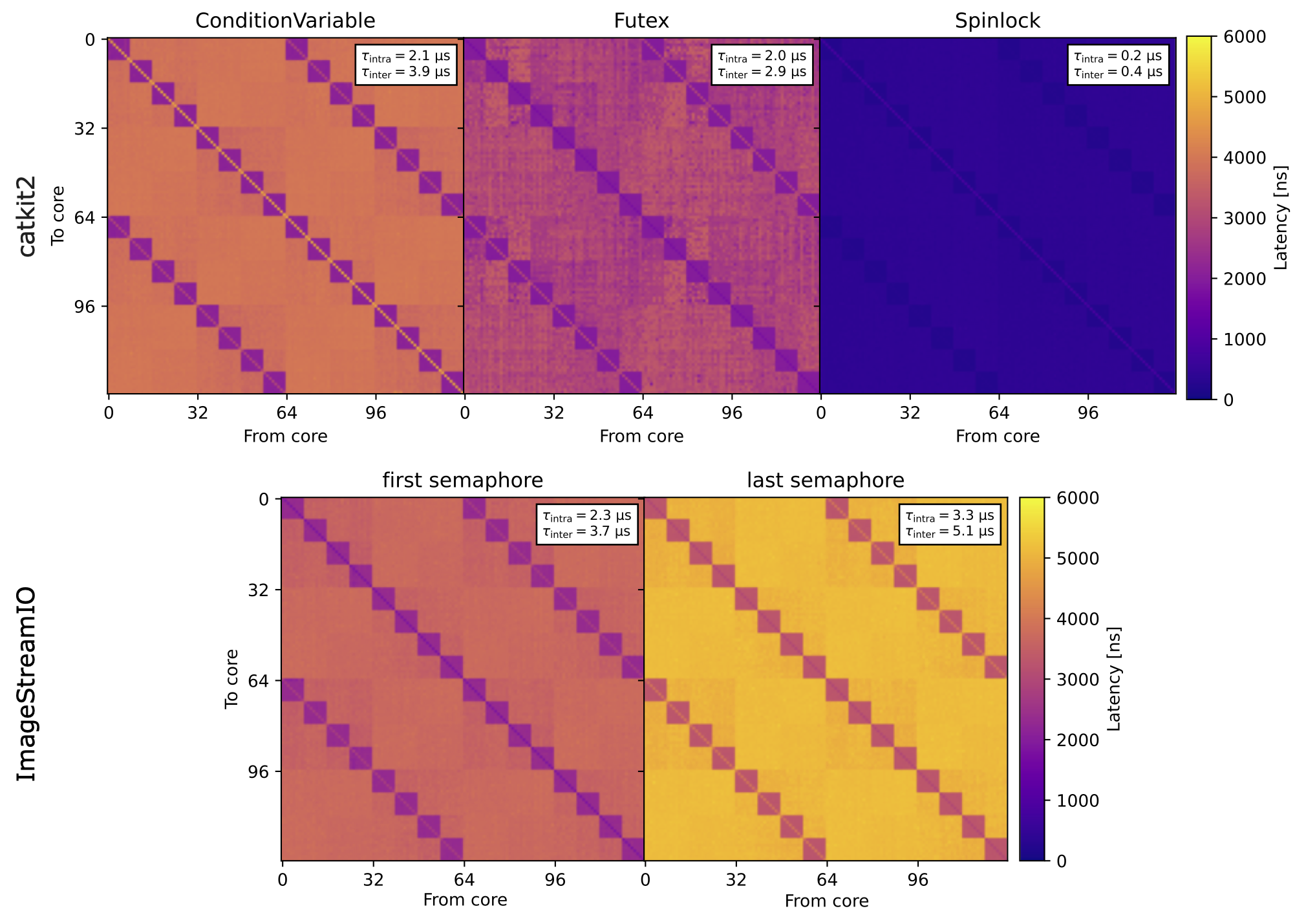}
    \caption{The core-to-core latency for some of the implemented wait methods in Catkit2, and comparison with the CACAO semaphore implementation when waiting either on the first and the last of the semaphores. Benchmarks were performed on Linux kernel 5.15.0-139 without real-time patch, running on an AMD Ryzen Threadripper PRO 5995WX 64-core CPU. The chiplet architecture and hyperthreading mode of this CPU is clearly visible, with 8 cores and 16 threads per chiplet. We list the mean inter and intra-chiplet latencies for each panel. Under these test conditions, Catkit2 demonstrates improved latency performance compared to ImageStreamIO, when using its futex implementation. If the utmost latency performance is required, Catkit2 provides spin-lock waits that yield sub-microsecond latencies at the cost of wasting CPU cycles.}
    \label{fig:core_to_core_latency_comparison}
\end{figure}

To address this need, Catkit2 has implemented a high-speed message-passing interface for inter-service communication. While conceptually related to the messaging system using in the SPIDERS instrument \cite{thompson2024rtc}, our performance requirements led to the development of a custom message broker. Local services exchange data through a single shared-memory message buffer. Memory for messages is dynamically allocated using a lock-free buddy allocator \cite{marotta2018non}. This ensures that no service can block another service from submitting a message. Notification of new message publications is provided by futexes, semaphores, condition variables, or spinlocks depending on the operating system and user choice. This enables latencies from a few microseconds up to sub-microsecond between the publishing of a message and access to that message on another process running on the same node. A measurement of core-to-core latencies is shown in Figure~\ref{fig:core_to_core_latency_comparison}. Inter-node communication is under development. ZeroMQ will be used to synchronize and transfer messages between message brokers running on the different nodes of SEAL. At present, a select number of services use dedicated ZeroMQ connections for operations.

\begin{figure}[t]
    \centering
    \includegraphics[width=0.9\columnwidth]{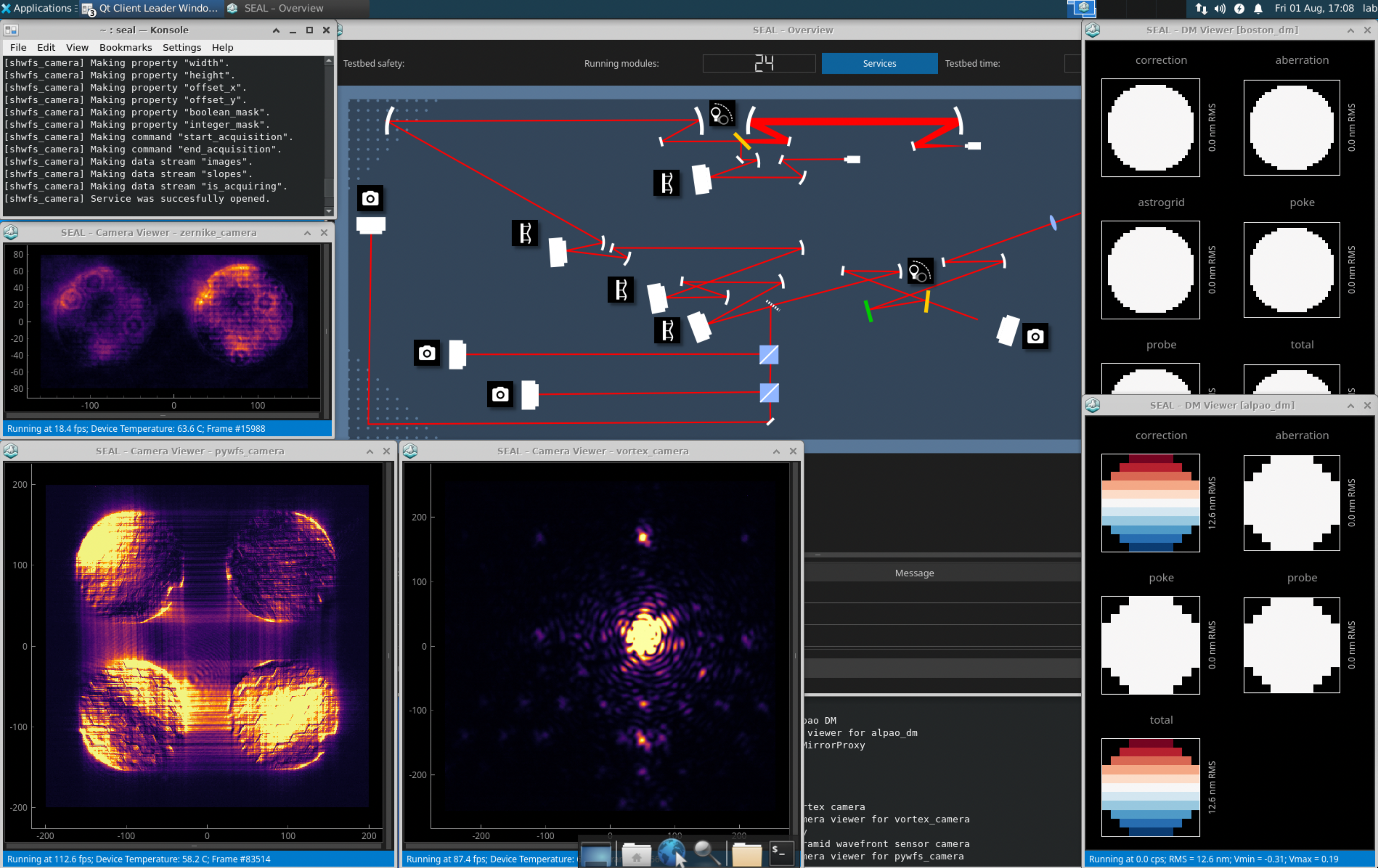}
    \caption{The graphical user interface of the Catkit2 RTC on the SEAL testbed. On the left the camera images for the vector-Zernike WFS and the Pyramid WFS. On the bottom a non-coronagraphic image in the vortex coronagraph branch. The vortex coronagraph mask was removed at the time this screenshot was taken. On the right the current DM commands on our Boston Micromachines MEMS tweeter and ALPAO woofer DMs.}
    \label{fig:user_interface}
\end{figure}

A graphical user interface has also been developed to support SEAL operations. The current implementation provides monitoring of deformable mirror commands, camera images, and overall testbed status. A representative screenshot is shown in Figure~\ref{fig:user_interface}. Future extensions will enable control of calibration routines and feedback loops through this interface.

\subsection{CACAO as a common real-time control across many platforms}
CACAO\footnote{\url{https://github.com/cacao-org/cacao}} (Compute And Control for Adaptive Optics) is a mature, flexible, and modular open-source software package, originally developed for the SCExAO instrument on the Subaru Telescope. It is designed to exploit available CPUs and GPUs efficiently, while managing heterogeneous data streams (wavefront sensors, cameras, telemetry) operating at different frame rates. Written primarily in C for execution speed, CACAO is well suited for real-time adaptive optics algorithms.

Detailed descriptions of CACAO are presented in several proceedings \cite{cacao2018, Guyoncacao2020, skaf2025}. Because many RTCs worldwide have adopted aspects of CACAO, most notably its shared-memory structure, or were inspired by it, this motivated the choice to instal CACAO on SEAL. This enables greater flexibility for collaboration across teams and facilitates smoother on-sky deployment. Importantly, CACAO is designed as a development RTC, making it ideal for wavefront sensing and control development.

The previous SEAL RTC was kRTC / DAO \cite{Cetre2018}, an offshoot of CACAO. However, its shared-memory format differed from CACAO’s. By implementing CACAO on SEAL, kRTC and DAO could be transitioned to a common shared-memory format. Once this was extended to all bench hardware, software–hardware interactions became significantly more straightforward. With this infrastructure in place, the AO loop was successfully closed using the pyramid wavefront sensor at $\sim$15 Hz.

\section{CURRENT PROJECTS AND FUTURE WORK}
\label{sec:current}

\noindent \textbf{Photonic Coronagraphy:} Photonic coronagraphy is a new approach to high-contrast imaging that replaces bulk optics with photonic integrated circuits (PICs) \cite{sirbu2024astropic}. In this concept, light from the telescope is injected into a PIC, where it propagates through a mesh of Mach–Zehnder interferometers (MZIs). By carefully tuning the thermal phase shifters in the mesh, the starlight is directed into a small number of output waveguides, while the remaining waveguides are nulled and primarily carry planet light. This mode separation enables the PIC to act as a coronagraph. With a different configuration, the same device can also function as a wavefront sensor, making use of the separated starlight channels for calibration and control. The AstroPIC program is currently developing and testing this concept with pupil-plane injection into the PIC. Our complementary work on the SEAL testbed focuses instead on focal-plane injection, opening up a different operational regime. Injection into the PIC is achieved with a microlens array aligned to grating couplers on the chip surface. Two focal-plane injection configurations will be tested. In the first, unfiltered focal-plane light is injected directly into the PIC, leaving the chip to perform the coronagraphic mode separation by itself. In the second, light is first processed by a bulk-optic phase-apodized pupil Lyot coronagraph (PAPLC) \cite{por2020paplc} and then injected into the PIC, forming a hybrid photonic coronagraph architecture \cite{desai2023photonic}. This latter configuration is expected to improve robustness against PIC manufacturing tolerances and stray light at minimal loss of planet throughput.

\noindent \textbf{Photonic Lantern WFS:} A photonic lantern (PL) was temporarily installed inside the SEAL testbed in 2023 in order to support the preliminary closed-loop results presented in Sengupta et al.~2024\cite{sengupta2024}. This work highlighted the need for a simplified, all-IR testbed for PL wavefront sensing development, which led us to build the \textbf{m}iniat\textbf{u}re \textbf{i}nfra\textbf{r}ed Santa cruz Extreme AO Laboratory (muirSEAL). The muirSEAL testbed, described in Sengupta et al.~2025 in these proceedings\cite{Sengupta25}, is inside SEAL's enclosure but otherwise separate from SEAL. It is currently being used to explore the relationship between the f-number of the beam incident on the PL and the PL's wavefront sensing performance (Sengupta et al.~2025 in these proceedings\cite{Sengupta25}), the use of the PL for reconstructing segment co-phasing errors (Cuevas et al.~2025 in these proceedings\cite{Cuevas25}), and to support hardware and software development for the PL deployed at Lick Observatory\cite{deMartino2022,deMartino2024}. 

\noindent \textbf{Zernike WFS:} The focus of the Zernike WFS work on SEAL is to support technology development for the co-phasing of segmented primary mirrors, in particular for WMKO and the Habitable Worlds Observatory (HWO). The next step will be to address the question of how to disentangle primary mirror co-phasing errors from static or quasi-static aberrations that originate downstream of the primary -- a long-standing challenge at WMKO. We will explore the relationship between Zernike mask design parameters (such as the dimple diameter) and our sensitivity to the mid-spatial scale frequency regime that is relevant to segment co-phasing. Additionally, we will compare the impact of different Zernike WFS reconstructors and control algorithms on SEAL's final contrast ratio using the vector vortex coronagraph. This work is further described in Salama et al.~2025 in these proceedings and Salama et al.~in prep HWO25 Proceedings. 

\noindent \textbf{Predictive WFC:} Predictive wavefront control describes a collection of methods that apply wavefront corrections based on estimations of future wavefronts (for a review see Fowler \& Landman 2023\cite{Fowler23}), with the goal of mitigating errors induced by the time-lag of the AO system. The SEAL testbed is currently being used to compare data and model driven prediction (Empirical Orthogonal Functions \cite{Guyon17} and Predictive Fourier Control \cite{Poyneer07} respectively) to inform current 8-10 meter class observatories as well as future ELTs. This project started with a Keck-like simulation, comparing the two predictors on a simulated Maunakea atmosphere and Keck-II AO telemetry\cite{Fowler22}. The SEAL testbed acts as a natural next step to continue this comparison in support of future integration of predictive methods into the Keck AO system. Initial work to implement EOF on the refractive SEAL testbed is described in Fowler et al.~2024\cite{Fowler24}. To date, the precursor work needed to implement a Shack-Hartmann WFS interface in Catkit2 has been completed. We plan to compare data and model driven predictive methods on the reflective rebuild, using the Shack-Hartmann WFS, the ALPAO deformable mirror as a corrector, and the Boston deformable mirror as a turbulence simulator. 

\acknowledgments % equivalent to \section*{ACKNOWLEDGMENTS}       
This work was supported by the National Science Foundation 
Advanced Technologies and Instrumentation for the Astronomical Sciences grant 2008822, The Heising-Simons Grants 2020-1822 and 2021-2373, and the University of California Observatories. EHP acknowledges support from the Heising-Simons Foundation through the 51 Pegasi b Fellowship.

% References
\bibliography{report} % bibliography data in report.bib
\bibliographystyle{spiebib} % makes bibtex use spiebib.bst

\end{document}